\documentclass
[preprint,twocolumn,prl,tightenlines,10pt,amsfonts,amssymb,balancelastpage,showpacs]{revtex4}%
\usepackage{amsfonts}
\usepackage{amsmath}
\usepackage{amssymb}
\usepackage{graphicx}%
\begin{document}
\title[Squeezing frequency combs]{Squeezing frequency combs}
\author{G. J. de Valc\'{a}rcel, G. Patera, N. Treps, C. Fabre}
\affiliation{Laboratoire Kastler Brossel, Universit\'{e} Pierre et
Marie Curie-Paris6, 4 place Jussieu CC 74, 75252 Paris cedex 05,
France}
\begin{abstract}
We have developed the full theory of a synchronously pumped type I
optical parametric oscillator (SPOPO). We derive expressions for
the oscillation threshold and the characteristics of the generated
mode-locked signal beam. We calculate the output quantum
fluctuations of the device, and find that, in the degenerate case
(coincident signal and idler set of frequencies), perfect
squeezing is obtained when one approaches threshold from below for
a well defined "super-mode', or frequency comb, consisting of a
coherent linear superposition of signal modes of different
frequencies which are resonant in the cavity.
\end{abstract}
\date{\today}
 \maketitle

Optical Parametric Oscillators are among the best sources of
squeezed \cite{squeeze}, correlated \cite{correlated} and
entangled \cite{Peng} light in the so-called continuous variable
regime. They have allowed physicists to successfully implement
demonstration experiments for high sensitivity optical
measurements and quantum information protocols. In order to
maximize the quantum effects, one needs to optimize the parametric
down-conversion process. This has been achieved so far by using
either intense pump lasers or resonant cavities. Having in mind
that the parametric process is an almost instantaneous one,
femtosecond mode-locked lasers are the best pump sources in this
respect, as they generate very high peak optical powers with high
coherence properties. Furthermore, they minimize the thermal
effects in the linear crystal which often hamper the normal
operation of parametric devices. Mode-locked lasers have been
already used extensively to generate non classical light, either
to pump a parametric crystal \cite{Grangier} or an optical fiber
\cite{Shelby}. However in such single-path configurations, perfect
quantum properties are only obtained when the pump power goes to
infinity. This is the reason why mode-locking is often associated
to Q-switching and pulse amplification \cite{mode lock Q switch}
in order to reach even higher peak powers, at the expense of a
loss in the coherence properties between the successive pump
pulses. In contrast, intracavity devices produce perfect quantum
properties for a finite power, namely the oscillation threshold of
the device. It is therefore tempting to consider devices in which
one takes advantage of the beneficial effects of both high peak
powers and resonant cavity build-up. Such devices exist: they are
the so-called synchronously pumped OPOs or SPOPOs. In a SPOPO, the
cavity round-trip time is equal to the repetition rate of the
mode-locked laser, so that the effect of the successive intense
pump pulses add coherently, thus reducing considerably its
oscillation threshold. Such SPOPOs have already been implemented
as efficient sources of tunable ultra-short pulses
\cite{manipSPOPO0,manipSPOPO1,manipSPOPO2,manipSPOPO3,manipSPOPO4,manipSPOPO5}
and their temporal properties have been theoretically investigated
\cite{theorySPOPO1,theorySPOPO3,theorySPOPO4}. To the best of our
knowledge, they have never been used so far to generate genuine
quantum effects. Performing a quantum analysis of this device, we
theoretically show in this paper that perfect squeezing can indeed
be obtained in SPOPOs as in cw OPOs. The squeezed mode is not a
usual single frequency mode, but instead a "super-mode", which is
a well defined linear combination of signal modes of different
frequencies, forming the frequency comb that will oscillate above
the SPOPO oscillation threshold.

\begin{figure}
\centerline{\includegraphics[width=.9\columnwidth]{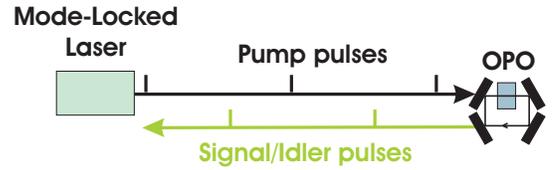}}
\caption{\label{figure} Synchronously pumped OPO}
\end{figure}

Let us first precise the model that we use (figure \ref{figure}).
We consider a ring cavity of optical length $L$ containing a type
I parametric crystal of thickness $l$. Degenerate phase matching
is assumed, what means that the phase-matching condition is
fulfilled for frequencies $2\omega_{0}$ and $\omega_{0}$. This
amounts to saying that $n\left( 2\omega_{0}\right) =n\left(
\omega_{0}\right) \equiv n_{0}$, $n\left( \omega\right) $ being
the refractive index of the crystal at frequency $\omega$. The
mode-locked pump laser, having a repetition rate
$\Omega/2\pi=c/L$, is tuned so that the frequency of one of its
modes is equal to $2\omega_{0}$. The electric field generated by
the pump mode-locked laser can be expressed as:
\begin{equation}
E_{\mathrm{ext}}\left( t\right) =\left( \frac{P}{2\varepsilon_{0}c}\right)
^{\frac{1}{2}}\sum_{m}i\alpha_{m}e^{-i\left( 2\omega_{0}+m\Omega\right)
t}+\mathrm{c.c.},\label{Eext}%
\end{equation}
where $P$ is the average laser power per unit area, $\alpha_{m}$
the normalized ($\sum\nolimits_{m}\left\vert \alpha_{m}\right\vert
^{2}=1$) complex spectral component of longitudinal mode labelled
by the integer index $m$, and $m=0$ corresponds to the
phase-matched mode. For the sake of simplicity in this first
approach to the problem, we will take the modal coefficients
$\alpha_{m}$ as real numbers, thus excluding chirped pump pulses.
As already mentioned, the SPOPO cavity length is adjusted so that
its free spectral range coincides with that of the pumping laser.
In the nonlinear crystal, pump photons belonging to all the
different longitudinal pump modes are converted into signal and
idler photons via the parametric interaction. In addition we will
assume here that we are in the ideal case of \textit{doubly
resonant degenerate operation}, meaning that the among all the OPO
cavity resonant frequencies, they are all the pump mode
frequencies $\omega _{p,m}=2\omega_{0}+ m \Omega$ but also all the
frequencies $\omega _{s,q}=\omega_{0} + q \Omega$ around the
phase-matched subharmonic frequency $\omega_{0}$. The intracavity
electric field generated by the parametric interaction will then
be a superposition of fields oscillating at frequencies
$\omega_{s,q}$. We will finally call $\gamma_{p}$ and
$\gamma_{s}$, the cavity damping rates for the pump and signal
modes. Note that the free spectral range $\Omega$ is assumed to be
the same in the pump and in the signal spectral regions. This is
necessary for an efficient intracavity parametric down conversion
and requires, from the experimental viewpoint, the use of extra
dispersive elements inside the cavity that compensate for the
dispersion of the crystal. At the quantum level, the signal field,
taken at the middle of the crystal, is represented by the quantum
operator $\hat{E}_{s}$ which can be written as:
\begin{equation}
\hat{E}_{s}(t)=\sum\nolimits_{q}i\mathcal{E}_{s,q}\hat{s}_{q}(t)e^{-i\omega
_{s,q}t}+\mathrm{H.c.},\label{Ea}%
\end{equation}
where $\hat{s}_{q}$ are the annihilation operators for the
$q^{th}$ signal mode in the interaction picture.
$\mathcal{E}_{s,q}$ is the single photon field amplitude, equal to
$\sqrt{\hbar\omega_{s,q}/2\varepsilon_{0}n\left(
\omega_{s,q}\right) AL}$, and $A$ its effective transverse area.
The following Heisenberg equations for the field operators can be derived
using the standard methods. The detail of the derivation will be given in a
forthcoming publication. Below threshold, and in the linearized regime for the
pump fluctuations, they read:
\begin{equation}
\frac{d\hat{s}_{m}}{dt}=-\gamma_{s}\hat{s}_{m}+\gamma_{s}\sigma\sum
\nolimits_{q}\mathcal{L}_{m,q}\hat{s}_{q}^{\dag}+\sqrt{2\gamma_{s}}\hat
{s}_{\mathrm{in},m},\label{d1}%
\end{equation}
where $\sigma$ is the normalized pump amplitude
\begin{equation}
\sigma=\sqrt{P/P_{0}}%
\end{equation}
in which $P_{0}$ is the single mode c.w. oscillation threshold:
\begin{equation}
P_{0}=2\gamma_{s}^{2}\gamma_{p}n_{0}^{3}c^{3}\varepsilon_{0}/\left( 4\sqrt
{2}\chi l\omega_{0}\right) ^{2}%
\end{equation}
with $\chi$ the crystal nonlinear susceptibility. $\mathcal{L}_{m,q}$ is the
product of a phase-mismatch factor by the pump spectral normalized amplitude
$\alpha_{m+q}$:
\begin{equation}
\mathcal{L}_{m,q} =\frac{\sin\phi_{m,q}}{\phi_{m,q}} \alpha_{m+q},\label{Lmq}%
\end{equation}

The phase mismatch angle
\begin{equation}
\phi_{m,q}=\frac{l}{2}\left( k_{p,m+q}-k_{s,m}-k_{s,q}\right)
\end{equation}
can be computed using a Taylor expansion around $2\omega_{0}$ for the pump
wave vectors $k_{p,m}$ and around $\omega_{0}$ for the signal wave vectors
$k_{s,q}$:
\begin{equation}
\phi_{m,q}\simeq\beta_{1}\left( m+q\right) +\beta_{2p}\left( m+q\right)
^{2}-\beta_{2s}\left( m^{2}+q^{2}\right) ,\label{fiqm}%
\end{equation}
where $\beta_{1}=\frac{1}{2}\Omega\left(
k_{p}^{\prime}-k_{s}^{\prime }\right) l$,
$\beta_{2p}=\frac{1}{4}\Omega^{2}k_{p}^{\prime\prime}l$,
$\beta_{2s}=\frac{1}{4}\Omega^{2}k_{s}^{\prime\prime}l$.
$k^{\prime }$ and $k^{\prime\prime }$ are the first and second
derivative of the wave vector with respect to frequency. Finally
$\hat {s}_{\mathrm{in},m}$ are the input signal field operators at
frequency $\omega_{s,m}$ transmitted through the coupling mirror.
When the input is the vacuum state, which we consider here, their
only non-null correlations are:
\begin{equation}
\left\langle \hat{s}_{\mathrm{in},m_{1}}\left( t_{1}\right) \hat
{s}_{\mathrm{in},m_{2}}^{\dag}\left( t_{2}\right) \right\rangle
=\delta_{m_{1},m_{2}}\delta\left( t_{1}-t_{2}\right) .\label{corr-in}%
\end{equation}
In order to get Eqs. (\ref{d1}), we had to assume that $\mathcal{E}%
_{s,m}\simeq\mathcal{E}_{s,0}$ for all $m$ and to neglect the dispersion of
the nonlinear susceptibility. Both approximations require that pulses are not
too short. In usual practical conditions, pulse durations should be longer
than 20-30 \textrm{fs}.
Let us first determine the average values of the generated fields.
They are determined by the "classical" counterpart of Eq.
(\ref{d1}), removing the input noise terms, and replacing the
operators by complex numbers. The solution of these equations is
of the form $s_{m}\left( t\right) =S_{k,m}e^{\lambda_{k}t}$, where
$k$ is an index labelling the different solutions. The parameters
$S_{k,m}$ and $\lambda_{k}$ obey the following eigenvalue
equation:
\begin{equation}
\lambda_{k}S_{k,m}=-\gamma_{s}S_{k,m}+\gamma_{s}\sigma\sum\nolimits_{q}%
\mathcal{L}_{m,q}S_{k,q}^{\ast}.\label{LkSkm}%
\end{equation}
As matrix $\mathcal{L}$ is both self-adjoint and real ($\mathcal{L}%
_{m,q}=\mathcal{L}_{q,m}$ real, see Eqs. (\ref{Lmq})--(\ref{fiqm})), its
eigenvalues $\Lambda_{k}$ and eigenvectors $\vec{L}_{k}$, of components
$L_{k,m}$, are all real. As $\gamma_{s}$ and $\sigma$ are also real, there
exist two sets of solutions of Eqs. (\ref{LkSkm}), that we will call
$S_{k,m}^{\left( +\right) }$ and $S_{k,m}^{\left( -\right) }$. The first
set is given by $S_{k,m}^{\left( +\right) }=L_{k,m}$ and the second one is
$S_{k,m}^{\left( -\right) }=iL_{k,m}$, with corresponding eigenvalues:
\begin{equation}
\lambda_{k}^{\left( \pm\right) }=\gamma_{s}\left(
-1\pm\sigma\Lambda _{k}\right) ,\label{lambda}
\end{equation}
Let us now label by index $k=0$ the solution of maximum value of
$|\Lambda _{k}|$: $|\Lambda_0|=\max\left\{ |\Lambda
_{k}|\right\}$.
When $\sigma\left\vert \Lambda_{0}\right\vert$ is smaller than 1, all the rates $\lambda_{k}%
^{\pm}$ are negative, which implies that the null solution for the
steady state signal field is stable. For the simplicity of
notations, we will take $\Lambda_0$ positive in the
following\cite{lam0}. The SPOPO reaches its oscillation threshold
when $\sigma$ takes the value $1/ \Lambda _{0} $, i.e. for a pump
power $P=P_{\mathrm{thr}}$ equal to:
\begin{equation}
P_{\mathrm{thr}}=P_{0}/\Lambda_{0}^{2},\label{Pthr}%
\end{equation}
We can now define the normalized amplitude pumping rate $r$ by
$r=\sigma \Lambda_{0}$, so that threshold occurs at $r=1$. We will
call eigen-spectrum the set of $S_{k,m}$ values for a given $k$,
which corresponds physically to the different spectral components
of the signal field, and critical eigen-spectrum $S_{0,m}^{\left(
+\right) }$, the one associated with $\lambda_{0}^{\left( +\right)
}$, which changes sign at threshold. Above threshold, this
critical mode will be the "lasing" one, i.e. the one having a
non-zero mean amplitude when $r>1$. Let us note that the
eigen-spectrum in quadrature with respect to the critical one,
$S_{0}^{\left( -\right) }=iS_{0}^{\left( +\right) }$, has an
associated eigenvalue $\lambda _{0}^{\left( -\right)
}=-2\gamma_{s}$ at threshold. Furthermore, equation (\ref{lambda})
implies that all the damping rates $\lambda_{k}^{\left( \pm\right)
}$ are comprised below threshold between $-2\gamma_{s}$ and $0$,
and that, whatever the pump intensity, all the eigenvalues
$\lambda_{k}^{\left( \pm\right) }\left( r\right) $ lie between
$\lambda_{0}^{\left( +\right) }\left( r\right) $ and
$\lambda_{0}^{\left( -\right) }\left( r\right) $. These properties
will be useful for the study of squeezing. To determine the
threshold, we must find $\Lambda_{0}$, and therefore diagonalize
$\mathcal{L}$, which cannot be done analytically in the general
case. The detailed study of the SPOPO threshold will be made in a
forthcoming publication. Here we will consider a special case that
leads to simple calculations and corresponds to an optimized
situation. We assume first that $\beta_{1}=0$ (equal group
velocities) and $\beta_{2s}=2\beta_{2p}$ (matched group velocity
dispersions). We take simple forms for the phase matching
coefficient
\begin{equation}
\mathcal{L}_{mq}=e^{-\frac{1}{2}\eta\left\vert
\phi_{m,q}\right\vert} \alpha_{m+q}
\end{equation}
and for the normalized pump spectrum:
\begin{equation}
\alpha_{m}=\pi^{-1/4}\Delta_{p}^{-1/2}e^{-\frac{m^2}{2 \Delta _{p}
^2}}.
\end{equation}
If in addition we impose the relation
\begin{equation}
\eta\left\vert \beta_{2p}\right\vert =\Delta_{p}^{-2},
\end{equation}
giving comparable widths for the phase matching curve and the
laser spectrum, we can analytically diagonalize the matrix
$\mathcal{L}_{m,q}$. We find that its (unnormalized) eigenvectors
have the following components:
\begin{equation}\label{L}
L_{k,m}=e^{-\frac{m^2}{2 \Delta _{p} ^2}} H_{k}\left(
m/\Delta_{p}\right) ,
\end{equation}
$H_{k}$ being the Hermite polynomial of order $k$, and that its eigenvalues
are:
\begin{equation}
\Lambda_{k}=\delta_{k,0}\quad
\pi^{1/4}\sqrt{\frac{\Delta_{p}}{2}},
\end{equation}
This gives finally a SPOPO threshold (Eq. (\ref{Pthr})) equal to:
\begin{equation}
P_{\mathrm{thr}}=\frac{2
P_{0}}{\sqrt{\pi}\Delta_{p}},\label{SPOPthr}
\end{equation}
the associated eigenspectrum $S_{0,m}^{(+)}$ being a Gaussian of
width $\Delta_{p}$. Relation (\ref{SPOPthr}) means that the usual
c.w. OPO oscillation threshold is divided by a quantity roughly
equal to the number of pump modes, at least in the fully optimized
SPOPO that we have considered here. This has a simple explanation:
since the parametric coupling is very fast, the OPO actually
oscillates when the instantaneous peak power of the pump exceeds
the c.w. threshold. Having in mind that $P_{0}$ is on the order of
a few tens of $mW$ and that the number of
coherently oscillating modes in a mode-locked laser can easily exceed $10^{4}%
$, one can then envision SPOPO thresholds of the order of a
microWatt. Experimental implementations of such SPOPOs have not
reached such ultra-low thresholds because of the difficulty to
fulfill simultaneously, in a real experiment, all these precise
conditions.
We can now determine the squeezing properties of the signal field
in a SPOPO below threshold. This is done by using the SPOPO
linearized quantum equations. Let us introduce the operator
$\hat{S}_{in,k}(t)$ by:
\begin{equation}\label{Sk}
\hat{S}_{in,k}(t)=\sum\nolimits_{m}L_{k,m}\hat{s}_{in,m}(t)
\end{equation}
Because $\sum\nolimits_{m}|L_{k,m}|^{2}=1$, one has $\left[
{S}_{in,k}(t),{S}_{in,k}^{\dagger}(t^{\prime})\right]
=\delta(t-t^{\prime})$: $\hat{S}_{in,k}$ is the annihilation
operator of a combination of modes of different frequencies, which
are the eigen-modes of the linearized evolution equation
(\ref{d1}). The corresponding creation operator applied to vacuum
state creates a photon in a single mode, which can be labelled as
"super-mode", which globally describes a frequency comb. Defining in the same
way as in (\ref{Sk}) the intracavity operator $\hat{S}_{k}(t)$,
one can then write:
\begin{equation}
\frac{d}{dt}\hat{S}_{k}=-\gamma_{s}\hat{S}_{k}+\gamma_{s}\sigma\Lambda_{k}%
\hat{S}_{k}^{\dagger}+\sqrt{2\gamma_{s}}\hat{S}_{\mathrm{in},k},\label{d3}%
\end{equation}
Let us now define quadrature hermitian operators
$\hat{S}_{k}^{\left( \pm\right) }$ by:
\begin{align}
\hat{S}_{k}^{\left( +\right) } & =\hat{S}_{k}+\hat{S}_{k}^{\dagger}\\
\hat{S}_{k}^{\left( -\right) } & =-i\left( \hat{S}_{k}-\hat{S}%
_{k}^{\dagger}\right)
\end{align}
which obey the following equations:
\begin{equation}
\frac{d}{dt}\hat{S}_{k}^{\left( \pm\right) }=\lambda_{k}^{\left(
\pm\right) }\hat{S}_{k}^{\left( \pm\right) }+\sqrt{2\gamma_{s}}\hat
{S}_{\mathrm{in},k}^{\left( \pm\right) },\label{d4}%
\end{equation}
with $\lambda_{k}^{\left( \pm\right) }$ given by Eq.
(\ref{lambda}). These relations enable us to determine the
intracavity quadrature operators in the Fourier domain
$\tilde{S}_{k}^{\left( \pm\right) }(\omega)$
\begin{equation}
i\omega\tilde{S}_{k}^{\left( \pm\right) }(\omega)=\lambda_{k}^{\left(
\pm\right) }\tilde{S}^{\pm}(\omega)+\sqrt{2\gamma_{s}}\tilde{S}%
_{\mathrm{in},k}^{\left( \pm\right) }(\omega).\label{d2}%
\end{equation}

Finally, using the usual input-output relation on the coupling mirror:
\begin{equation}
\tilde{s}_{\mathrm{out},m}(\omega)=-\tilde{s}_{\mathrm{in},m}(\omega
)+\sqrt{2\gamma_{s}}\tilde{s}_{m}(\omega),\label{inout}%
\end{equation}
which extends by linearity to any super-mode operator because the mirror is
assumed to have a transmission independent of the mode frequency, one obtains
finally the following expression for the output signal super-mode quadrature
component in Fourier space:
\begin{equation}
\tilde{S}_{\mathrm{out},k}^{\left( \pm\right)
}(\omega)=\frac{\gamma _{s}\left( 1\pm r\Lambda_{k}/\Lambda_{0}
\right) -i\omega}{\gamma_{s}\left( -1\pm r\Lambda_{k}/ \Lambda
_{0}\right) +i\omega}\tilde{S}_{\mathrm{in},k}^{\left(
\pm\right) }(\omega),\label{d5}%
\end{equation}
These expressions are particularly simple for the critical mode quadrature
components ($k=0$):
\begin{equation}
\tilde{S}_{\mathrm{out},0}^{\left( \pm\right) }(\omega)=\frac{\gamma
_{s}\left( 1\pm r\right) -i\omega}{-\gamma_{s}\left( 1\mp r\right)
+i\omega}\tilde{S}_{\mathrm{in},0}^{+}(\omega),\label{d6}%
\end{equation}

The variance of these operators can be indeed measured using the usual
balanced homodyne detection scheme: the local oscillator is in the present
case a coherent mode-locked multimode field $E_{\mathrm{L}}\left( t\right) $
having the same repetition rate as the pump laser:
\begin{equation}
E_{\mathrm{L}}\left( t\right) =i\epsilon_{L} \sum_{m}
e_{m}e^{-i\omega_{s,m}t}+\mathrm{c.c.},
\end{equation}
where $\sum_{m}|e_{m}|^{2}=1$, and $\epsilon_{L}$ is the local
oscillator field total amplitude factor. Assuming that the
photodetectors measure the intensity of the Fourier components of
the photocurrent averaged over many successive pulses, the
balanced homodyne detection scheme measures the variance of the
fluctuations of the projection of the output field on the local
oscillator mode when the mean field generated by the OPO is zero,
which is the case below threshold. As a result, when the
coefficients $e_{m}$ of the local oscillator field spectral
decomposition are equal to the coefficients $L_{k,m}$ of the
$k$-th super-mode, one measures the two following variances,
depending on the local oscillator phase:
\begin{align}
V_{k}^{-}\left( \omega\right) &=\frac{\gamma_{s}^{2}\left(
1-r\Lambda_{k}/\Lambda_{0}\right)^{2}+\omega^{2}}{\gamma_{s}
^{2}\left( 1+r\Lambda_{k}/\Lambda_{0}\right) ^{2}+\omega^{2}}\label{VHgena}\\
V_{k}^{+}\left( \omega\right) &=V_{k}^{-}\left( \omega\right)
^{-1}\label{VHgenb}
\end{align}
Eqs. (\ref{VHgena},\ref{VHgenb}) shows that quantum noise
reduction below the standard quantum limit (equal here to 1) is
achieved for any super-mode characterized by a non-zero
$\Lambda_{k}$ value and that the smallest fluctuations are
obtained close to threshold and at zero Fourier frequency:
\begin{equation}
\left(V_{k}\right) _{\min}=\left( \frac{\Lambda_{0}-|\Lambda_{k}|
}{\Lambda_{0}+|\Lambda_{k}|}\right)^{2}\label{Squeeze}
\end{equation}
In particular, if one uses as the local oscillator the critical
mode $k=0$, identical to the one oscillating just above the
threshold $r=1$, one then gets perfect squeezing just below
threshold and at zero noise frequency, just like in the c.w.
single mode case. But modes of $k \neq 0$ may be also
significantly squeezed, provided that $|\Lambda_k/\Lambda_0|$ is
not much different from $1$. In conclusion, we have studied the
quantum behaviour of a degenerate synchronously-pumped OPO, which
seems at first sight a highly multimode system, since it involves
roughly $10^5$ different usual single frequency modes for a
$100fs$ pulse. We have shown that its properties are more easily
understood if one considers the "super-modes", linear combinations
of all these modes that are eigen-modes of the SPOPO set of
evolution equations and describe in a global way the frequency
comb -or, equivalently, the train of pulses- generated by the
SPOPO. The super-mode of minimum threshold plays a particular
role, as it is the one which turns out to be perfectly squeezed at
threshold and will oscillate above threshold. The present paper
gives a first example of the high interest of studying frequency
combs at the quantum level, as they merge the advantages of two
already well-known non-classical states of light: the c.w. light
beams, with their high degree of coherence and reproducibility and
the single pulses of light, with their high peak power enhancing
the non-linear effects necessary to produce pure quantum
effects.\\

Laboratoire Kastler Brossel, of the Ecole Normale Sup\'{e}rieure
and the Universit\'{e} Pierre et Marie Curie-Paris6, is UMR8552 of
the Centre National de la Recherche Scientifique. G.J. de V. has
been financially supported by grant PR2005-0246 of the
Secretar\'{\i}a de Estado de Universidades e Investigaci\'{o}n del
Ministerio de Educaci\'{o}n y Ciencia (Spain). Its permanent
address is: Departament d'\`{O}ptica, Universitat de Val\`{e}ncia,
Dr. Moliner 50, Val\`{e}ncia, Spain


\begin{thebibliography}{99}
\bibitem {lam0} Should $\Lambda_{0}<0$ then the null eigenvalue at threshold
would be $\lambda_{0}^{\left( -\right) }$ instead of
$\lambda_{0}^{\left( +\right) }$ and the following analysis
should be accordingly modified.
\end{thebibliography}
\end{document}